# Correlated "noise" in LIGO gravitational wave signals: an implication of Conformal Cyclic Cosmology

by Roger Penrose
Mathematical Institute, Oxford, UK

*Abstract*

It has recently been reported by Cresswell *et al*. [1] that correlations in the *noise* surrounding the observed gravitational wave signals, GW150194, GW151226, and GW170194 were found by the two LIGO detectors in Hanford and Livingston with the same time delay as the signals themselves. This raised some issues about the statistical reliability of the signals themselves, which led to much discussion, the current view appearing to support the contention that there is something unexplained that may be of genuine astrophysical interest [2]. In this note, it is pointed out that a resolution of this puzzle may be found in a proposal very recently put forward by the author [3], see also [4], that what seems to be spuriously generated noise may in fact be gravitational events caused by the decay of dark-matter particles (erebons) of mass around $10^{-5}$g, the existence of such events being a clear implication of the cosmological scheme of conformal cyclic cosmology, or CCC [5], [6]. A brief outline of the salient points of CCC is provided here, especially with regard to its prediction of erebons and their impulsive gravitational signals.

## 1. Dark matter decay, as an astrophysical source of impulsive gravitational signals

A dispute has recently arisen concerning correlations that have been pointed out between "noise" events in the data collected by the LIGO gravitational wave detectors at Hanford WA and Livingston LA. According to Cresswell *et al* [1], the time delay (6.9ms) between the reception of the gravitational wave signal GW150509 at these two locations was also the same time delay that provided a surprising correlation between the noise in the signal, occurring before, after, and during the actual wave form of the gravitational-wave signal— where that wave form is convincingly interpreted as coming from an inspiralling pair of black holes. A similar effect was observed also in the black-hole encounter events GW151226 and GW170104. This correlation is, on the face of it, hard to account for, since the concept of "noise" here, would normally be expected to be some local effect, in the neighbourhood of each of the two detectors, at Hanford and, Livingston, which are separated by some 3000km. Such correlation would appear to imply that a significant component of this "noise" has actually a cosmological or astrophysical origin, rather than a local one. If such an external explanation can be provided, then this could resolve the dispute referred to at the beginning of this section.

In this note, I point out that there is a clear astrophysical candidate for what could well be a major component of these "noise" signals [3]. I contend that they could, in large part, be actual gravitational signals, of an unusual type, these being clearly predicted to occur on the basis of a particular cosmological proposal, referred to as *conformal cyclic cosmology*, or CCC, which was originally put forward in 2005/6. Supporting observational evidence for CCC was subsequently published [6], [7], [8], [9], though disputed by others [10]. The unusual gravitational signals referred to above were recognized as being implications of that theory only very recently (see [3]). A brief outline of CCC, and its significant observational support will be given in Section 3. In this section, I simply assert the implications of CCC that are relevant to the issue at hand without providing justification from CCC or otherwise. Such justification will be provided in Section 4.

It is indeed a reasonably robust implication of CCC that dark matter consists of particles—that I refer to as *erebons* (after the ancient Greek God of Darkness *Erebos*) that have a *Planck mass*, or thereabouts ($\sim 10^{-5}g$), and a decay lifetime that could well be in excess of some $10^{10}$ years. Erebons are scalar particles (having zero spin), and interact only gravitationally (this much being a necessary feature of their key role in CCC as a kind of scalar conformal partner to gravitation). Because of its large mass, each erebon should behave effectively like a *classical* particle—according to various proposals that "self-measurement" should occur in quantum systems that involve sufficient mass displacement [11], [12], [13], [14], [15]. Moreover, its decay would, correspondingly, take the form of, a total conversion into a highly oscillatory *classical* gravitational wave, travelling outwards from the erebon's location, in an effectively classical impulsive spherical shell.

Since the oscillation frequency would be (initially, at least) of the absurdly large order of a Planck frequency ($\sim 10^{43}$ Hz), there is no way that the oscillatory nature of the wave would be picked up in any presently feasible gravitational wave detector. Instead, we have to consider that, in effect, the wave would behave more like an instantaneous impulse that would be of a confusing nature for those who are used to studying ordinary macroscopic gravitational waves. We think of such a macroscopic gravitational wave as having the effect, on any material that it encounters, as inflicting a *squashing* (inward acceleration) in one direction transverse to the direction of the wave, and a *stretching* (outward acceleration) in the perpendicular transverse direction. These effects are simply those due to the *Weyl*-tensor (i.e. conformal) part of the Riemann curvature tensor. For an oscillatory wave, this squashing/stretching would be followed by the opposite effect (stretching/squashing), and so on repeatedly. To lowest order, we would expect that, overall, these squashing and stretching effects would ultimately cancel out when the wave has passed, but there would be a higher-order residual for which there would be a transverse *squashing* in *all* directions transverse to the direction to the wave. The latter would resemble the gravitational effect of the passage of a *matter* wave (e.g. electromagnetic), as given by a delta function in the *Ricci* tensor. These effects can be easily understood by considerations of ordinary classical lens optics, see [16], especially p.266. For a brief Planck-frequency gravitational wave, the resultant would appear to be an impulsive matter wav, carrying away the mass-energy of the erebon. (See [17] for this concept of an impulsive wave.) The effective space-time geometry is that of what could be referred to as an "impulsive Vaidya metric" (which is a Schwarzschild space-time converting itself to Minkowski space-time along an outgoing spherically symmetric null hypersurface, which would be an expanding spherical shell of delta-function Ricci tensor.

## 2. Erebon decays as an astrophysical probe.

If these effects of CCC are respected by nature, they should provide the potential for a powerful new way of exploring the universe. Conversely, if the stringent implications of this idea turn out *not* to be respected by nature, then it should not be too hard to establish this fact, and therefore refute the proposal that I am putting forward here. Of course, it might also be a possibility that something of the general nature of that which I am describing is indeed true, but the model that I am describing needs modification from the basic picture being presented here. In any case, this proposal presents a possible resolution of the puzzle that appears to be posed by the correlations in the "noise" in the LIGO signals described at the beginning of this article.

Let us consider some relevant issues

(1) Since the erebon decays, are gravitational wave events, they should have an effect on a gravitational wave detector such as LIGO, though the signal is different from what is normally expected in the usual applications of Einstein theory, owing to the extremely high ($\approx 10^{43}$ Hz) frequency expected in the signal. Here the signal should be impulsive (instantaneous), with *inward* acceleration in all directions transverse to the direction of the wave.

(2) The signals would have a random character that would make them easily mistaken for irrelevant noise effects.

(3) If the theory is correct, these signals would provide a direct signal from dark matter concentrations, which could be compared with other estimates of where and how dense such distributions might be expected to be.

(4) Since it is supposed that all erebons have the same mass ($\approx 10^{-5}$g), where their decays provide point-like gravitational explosions of about the Planck energy ($\approx 2 \times 10^9$J $\approx 10^{19}$GeV) which is that of a sizable artillery shell, the strength of the received signal should provide a good measure of the distance of the source, if its radial velocity is otherwise known (or of its radial velocity if its distance is otherwise known).

(5) Any serious inconsistency with otherwise known facts would tell against the theory, so there is a considerable tightness in the overall scheme, making it extremely prone to observational disproof—which is a desirable feature of any serious scientific proposal.

(6) The above comments refer to what can be achieved by the use of data from the existing LIGO gravitational wave detectors. Of course more can be achieved when further detectors of the LIGO type are completed. Moreover, there is potentially a far greater scope for gravitational wave detectors based on BECs, as has been proposed by Ivette Fuentes and her collaborators [18-20], since, being very much smaller and far less expensive, could be produced in large numbers, giving a much better directional capacity than LIGO type detectors. Moreover, the BEC detectors would be more effective at high frequencies, and therefore better tuned to the impulsive waves here predicted from erebon decay.

The main uncertainty in the scheme as it stands appears to be the lifetime (or half-life) of erebon decay. We shall be seeing in Section 3 that, in CCC, there is a strong relation between this decay rate and the nature of the temperature fluctuations in our cosmic microwave background, which should, in principle give strong restrictions to what this decay can be, but this has not yet been looked at in any detail.

Let us now examine how erebon decays might influence the LIGO signals, appearing as "noise", but with the same time delay between the two detectors as have the black-hole encounter signals. The most obvious point is that the black hole pair would be expected to be in a galaxy, and that galaxy would, according to my proposal, be a source of erebon decay—where we bear in mind that the dark-matter density in a galaxy might be around a hundred times that in intergalactic space. Accordingly, that part of the "noise" which comes from the erebon decay from that particular galaxy would indeed be correlated with the same time delay between the two detectors as would the signal from the black-hole encounter event.

Of course, in this description, the black-hole encounter is almost irrelevant, serving merely to locate an individual galaxy in the sky. To test this, on need merely settle on some prominent galaxy—say the Andromeda galaxy—and look for correlations in the noise that has a time delay between the two detectors that corresponds to sources in the direction of that particular galaxy. The black-hole encounter would be completely unnecessary for this, and one does not have to wait for such occurrences in order to test the proposal. Of course, the dark-matter distribution in our own galaxy would be a major contributor, in this proposed scheme, and thus should be a major contributor to the signals referred to here. Clearly the proposal that I am putting forward here makes many testable predictions, and it should not be hard to disprove it if it is wrong.

## 3. The ideas of CCC

According to CCC—the conformal cyclic cosmology proposal—the universe consists of a (perhaps infinite) succession of *aeons*, where each aeon originates with its own big bang and has an unending exponentially expanding future, consistent with a positive cosmological constant $\Lambda$, taken to have the same constant value from aeon to aeon, the Einstein $\Lambda$-equations holding throughout every aeon. The driving idea behind CCC is the second law of thermodynamics (abbreviated 2$^{nd}$ Law) in the particular form that it appears to have been presented to us by nature, namely that the initial extremely low entropy at the beginning of each aeon is manifested in the fact that while the matter and radiation are thermalized immediately following the big bang of each aeon, the gravitational degrees of freedom are taken to be completely suppressed at its big bang, so that the Weyl tensor $C_{abcd}$ vanishes there. In CCC, this "Weyl curvature hypothesis" is replaced by K.P.Tod's proposal [21] that each aeon's big bang can be smoothly conformally extended backwards into another space-time region. That region is, according to CCC, taken to be the remote future (conformal infinity) of a previous aeon.

There is a theorem, due to H.Friedrich [22] that tells us that, with positive $\Lambda$ and electromagnetism as the matter content, a smooth spacelike conformal future infinity will be a generic situation, so that the conformal continuation of the remote future of one aeon to some other space-time is a generic possibility, ad according to CCC, it would be to the big bang of a further aeon.

As this stands, this is merely a plausible geometrical picture, and it may be difficult to take it seriously as a physically acceptable model. However, when examined from the physical point of view, it does indeed hang together. The physics near the big bang is at an enormous temperature, so rest mass is completely dominated by kinetic motions, and so may be ignored.

Moreover our current picture of the distant future of the universe, according to Einstein's $\Lambda$-equations is a continuing exponential expansion, with matter density continually reducing and an entropy content completely dominated by supermassive black holes. Eventually (after some $10^{100}$ years), even these supermassive black holes will have all evaporated away by Hawking radiation, to leave an exceedingly rarified universe whose particle number content will be almost entirely in the form of photons, so the conditions of Friedrichs's theorem are basically satisfied. There will, however, be a residue of a relatively small number of massive particles, and for the strict requirements of the theory one must apparently assume that there is a mechanism whereby in the hugely long-term future, mass eventually fades away, asymptotically—an admittedly unconventional hypothesis, but not contradicted by current observation, and not implausible on the basis of a particle physics which is extended from current ideas by the presence of a positive $\Lambda$.

Is it physically plausible that this enormously rarified and absurdly cold future infinity can be physically similar to the enormously dense and hot big bang of a succeeding aeon? Yes, indeed, if we may regard the relevant physics to be a conformally invariant one, where we bear in mind that not only are the equations of Maxwell conformally invariant, but so also are those of Yang-Mills, when mass can be ignored. This conformal freedom allows us to stretch out the hot big bang of the succeeding aeon and to squash down the remote future of the previous one—bearing in mind that energy and momentum scale in the exact inverse way to space and time. So hot becomes cold and dense becomes rarified upon conformal stretching.

A remark is necessary concerning the 2$^{nd}$ Law, as one might consider this problematic in a cyclic model. CCC's response is, first, to take note of the fact that, even now (according to the Bekenstein-Hawking entropy formula), by far the major contribution to the entropy in the universe is in supermassive black holes. This will vastly increase in the future. But ultimately, by Hawking radiation, all will disappear via Hawking evaporation. The issue here is that, in CCC, one must side with those who believe that information—or, more to the point, phase-space volume—must be destroyed by being obliterated at the hole-s singularity. As has often pointed out, this is in contradiction with quantum unitarity, but the view here is taken (in opposition to that of "many-worlds" proponents) that since unitarity is already violated in quantum measurement, then it is not a universal truth; moreover, the viewpoint will be taken here that in quantum measurement as well as with black-hole evaporation, quantum state reduction is a gravitational effect. This relates to the classicality of erebon particles, as already referred to in Section 1.

## 4. The equations of CCC

Of course one needs clear-cut equations to describe this transition from the remote future of one aeon to the big bang of the next. Such equations can be provided [6, Appendix A] [23], although there are still uncertainties, that will not concern us here, in certain detailed choices for ensuring uniqueness of evolution. Nevertheless, there are major issues concerning the evolution that are central to the current discussion.

The transition from each aeon to the next is described in terms of a "bandage metric" $g_{ab}$ that is defined throughout an open region $U$ containing the "crossover" hypersurface $X$ that connects the two aeons under consideration. Let $\hat{g}_{ab}$ be the physical metric for the aeon pre-crossover and $\check{g}_{ab}$, the physical metric post-crossover. We have conformal factors $\Omega$ and $\omega$ so that

$$\hat{g}_{ab} = \Omega^2 g_{ab} \quad \text{and} \quad \check{g}_{ab}, = \omega^2 g_{ab},$$

where the *reciprocal hypothesis*

$$\Omega = -\omega^{-1},$$

is taken to hold throughout U, and where ω is smooth, changing from negative to positive across X. Thus, Ω is infinite at X and represents a "squashing down" of the remote future of the earlier aeon so as to fit smoothly on to the "stretched out" big bang of the subsequent aeon across X within U provided by ω. Within the earlier aeon, the Ω-field has no *physical* content, but we can take it as a kind of auxiliary field—a kind of "phantom field"—which allows us to express Einstein's equations in a conformally invariant framework, and when taken alongside the gravitational field allows us to extend the range of Einstein's theory so that it includes the crossover 3-surface X. There are no physical degrees of freedom in Ω, beyond those already in the gravitational field. Yet it has an "energy tensor"

$$T_{ab}[\Omega] = \frac{1}{4\pi G}\Omega^3\{\nabla_a\nabla_b - \tfrac{1}{4}g_{ab}\nabla^c\nabla_c - \tfrac{1}{2}R_{ab} + \tfrac{1}{8}Rg_{ab}\}\Omega^{-1},$$

which is trace-free and divergence free and the Einstein Λ-equations become simply $T_{ab}=T_{ab}[\Omega]$, where $T_{ab}$ is the energy tensor of the matter (assumed to be massless).

However, on the future side of X, when the $\breve{g}_{ab}$ metric is used, the Ω behaves as an *actual* physical field (this being a consistent viewpoint because of the reciprocal hypothesis). This scalar field starts off as effectively massless, but would rapidly acquire a mass and would be expected to dominate over all other fields.

It is a hypothesis of CCC that this Ω-field, when it acquires a mass, is in fact the *dark matter* field that provides the dominant matter content of the universe (where Λ, or whatever else "dark energy" might be, does not count as matter). The gravitational degrees of freedom in the earlier aeon may be considered to be described by the Weyl tensor $\hat{C}_{abcd}$. We find that $\hat{C}_{abcd}$=0 at X (which is in accordance with the Weyl curvature hypothesis), but gravitational wave signals in that earlier aeon do register at X in the *normal derivative* of $\hat{C}_{abcd}$ across X. This information is not lost at crossover, but part of it (the "electric" part) is conveyed as higher derivatives in the Ω-field (the "magnetic" part being encoded into a a 3-space conformal curvature—the Cotton tensor—for X).

Such signals would affect the CMB of the succeeding aeon in observable ways. Up until now, interest has been concentrated on the prospect of observing, in our current aeon, the effect of supermassive black-hole encounters in the aeon prior to ours. The remote future of the emitted gravitational waves should, according to CCC, register in the CMB of our aeon, as circular features, sometimes in concentric sets (indicating several such events occurring in the same galactic cluster in the previous aeon), and claims have been made that signals are indeed seen in the WMAP data [6], [7], and also in the Planck data [8], [9], though disputes remain as to the significance of these evident signals [10]. It may be remarked that these observations demonstrate that the sources of these signals are highly inhomogeneously distributed across the sky, which is very problematic for the conventional explanation in terms of quantum fluctuations in the inflaton field.

Of more relevance to the present discussion are the almost scale-invariant temperature fluctuations in the CMB. This scale invariance is one of the strongest motivations for a belief in an exponentially expanding inflationary phase in our very early universe. In CCC, there cannot be any such inflationary phase in our aeon, but the role of inflation is taken over, instead, by the exponentially expanding ultimate history of the previous aeon. The source of these fluctuations cannot be inflatons, since there is no inflation in CCC. However, their role is taken over by decaying particles pervading the whole of space in the previous aeon, and these have to be the dark matter particles—the *erebons* referred to in earlier. Indeed these particles have ultimately to decay, in CCC, since we do not want them to build up from aeon to aeon. Their life-time has to be long, in comparison with the current age of our universe-aeon, so that the scale-invariance of the current exponential expansion, referred back to the aeon prior to ours can result in the observed close scale invariance in the CMB temperature fluctuations in our current aeon.

It is not the purpose of the current article to examine this feature in any detail. The concern, here, is to examine erebon decay in our own aeon, to see whether it might be an observable effect. The first point to take note of is the fact that no physical interactions other than gravity have been invoked in the above discussion. Erebons are scalar "conformal partners" to gravity. They would be expected to have no physical interactions other than gravitational. This is indeed consistent with their being the particles that constitute dark matter, since dark matter does actually appear to interact only gravitationally with other forms of matter. Since only gravity is involved, this suggests that the erebons should have a *Planck mass*, or a simple multiple of that mass, as put forward in Section 1. Finally, the fact that they should behave as classical particles (as proposed in Section 1), and, more importantly, as their decay should be in the form as a classical extremely high-frequency classical gravitational wave is crucial for the picture. A Planck-energy *quantum* particle would have a catastrophic effect on a detector which measured it, since the entire artillery-shell energy would be transferred to the detector upon measurement!

My thanks go to Iwo Bialynicki-Birula, Paul Tod, and most particularly to Krzysztof Meissner, discussions with whom led to the idea that dark-matter decay in the previous aeon might be responsible for the near scale invariant CMB temperature fluctuations, and to Ivette Fuentes, with whom discussions of possible detection of such decay in our own aeon using her proposed BEC detectors led me to the unusual line of thinking suggested here. I am also grateful for financial assistance through a personal endowment from J.P. Moussouris.